\documentstyle[aps,epsf,floats,twocolumn,prl]{revtex}
\begin{document}

\twocolumn[\hsize\textwidth\columnwidth\hsize\csname @twocolumnfalse\endcsname

\draft

\title{Dynamics and Transport in Random Antiferromagnetic
Spin Chains.}
\author{Kedar Damle$^{1,2}$, Olexei Motrunich$^1$ and David A. Huse$^1$}
\address{$^1$ Department of Physics, Princeton University, Princeton, NJ
08544 \\
$^2$ Department of Electrical Engineering, Princeton University, Princeton,
NJ 08544}
\date{November 9, 1999}
\maketitle
\begin{abstract}
{We present the first results on the low-frequency
dynamical and transport properties of random antiferromagnetic
spin chains at low
temperature ($T$). We obtain the momentum and
frequency dependent dynamic
structure factor in the Random Singlet (RS) phases of
both spin-1/2 and spin-1 chains, as well as in the
Random Dimer phase of spin-1/2 chains. We also show that the RS
phases are unusual `spin-metals' with divergent low-frequency
conductivity at $T=0$, and follow the
spin conductivity through `metal-insulator' transitions
tuned by the strength of
dimerization or Ising anisotropy in the spin-1/2 case, and by
the strength of disorder in the spin-1 case.}

\end{abstract}

]

Low-dimensional quantum spin systems in the presence of disorder are
a fascinating laboratory for the study of the interplay between quantum
effects, strong correlations and disorder.
This is in part because they are often very sensitive to disorder,
and display
dramatic effects due to small
amounts of disorder (especially in one dimension). The question of disorder
effects is thus of considerable experimental importance in such cases.
Moreover, a great deal is now known about the effects of
strong correlations and quantum fluctuations in pure one and
two-dimensional spin systems~\cite{review1},
allowing one to focus on the new features introduced by disorder. 

Random antiferromagnetic spin-1/2
chain compounds are interesting, experimentally realizable
examples~\cite{rafm1,rafm2} of such systems.
Theoretical work has led to the prediction that even a small
amount of disorder can cause these
systems to have an extremely unusual `Random Singlet' (RS)
ground state for a range of parameter values~\cite{dasma,doty,dsf}. In this RS
state, the interplay of disorder and quantum mechanics locks each
spin into a singlet pair with another spin; the two spins in a given
singlet pair can have arbitrarily large spatial separation.
A similar RS state is
also predicted in the spin-1 Heisenberg antiferromagnetic chain
for sufficiently strong randomness in the bonds~\cite{dsf,Monthus}.
The theory also yields
unusual results for the low-temperature thermodynamics and some
ground-state correlators in the RS states.

In this Letter, we report on the first theoretical study
of the dynamical and transport properties of these systems; the primary
motivation is of course to understand the dynamics and transport when
quantum interference, strong correlations and disorder are
all simultaneously important.
We focus here on the momentum and frequency dependent
dynamical structure factor $S(k,\omega)$ and the dynamical
spin conductivity. The former is directly probed by
inelastic neutron scattering (INS) experiments on these compounds, while the
latter is of considerable theoretical interest as it contains information
about the nature of any localization phenomena, or lack thereof, in
these strongly correlated disordered systems.

In particular, we calculate the low-temperature $S(k,\omega)$
at low frequencies in the RS states of spin-1/2 and spin-1 chains as well as
in the `Random Dimer' (RD)
phase of spin-1/2 chains that is predicted~\cite{yahybhgi} in the presence
of weak enforced bond-dimerization and strong disorder. Our results
lead us to expect some very unusual features in the INS cross-section as a
function of wavevector transfer $k$ (for fixed, small energy
transfer $\omega \agt T$), especially in
the RD phase.
For the conductivity, we calculate
the real part of the low-frequency dynamical spin
conductivity $\sigma^{\prime}(\omega)$ at low temperatures and
show that the RS phases of both spin-1/2 and spin-1 chains
are actually unusual `spin-metals' with a $\sigma^{\prime}(\omega)$
that {\em diverges} for low $\omega$ at $T=0$.
We also follow $\sigma^{\prime}(\omega)$
across novel `metal-insulator' transitions
to insulating states; these are accessed in the spin-1/2 case by turning on
dimerization (leading to a RD phase)
or by increasing the Ising anisotropy above a threshold, and in the
spin-1 Heisenberg case by reducing the disorder below a critical value.
Thus, we are able to obtain a wealth of {\em reliable} information about
the unusual dynamics and transport in these
strongly correlated random quantum systems. We emphasize that all our
results are exact in the low-frequency limit at $T=0$,
and continue to be valid for $T \not= 0$ so long as $T \alt \omega$ (in
some cases, we are also able to access frequencies $\omega < T$~\cite{us}).

In the spin-1/2 case, the specific problem we consider is the
random XXZ
Hamiltonian, which describes the low energy (magnetic) dynamics
of insulating antiferromagnetic spin-1/2 chain
compounds~\cite{rafm1,rafm2}
with chemical disorder that affects the bond strengths:
\begin{eqnarray}
{\cal H}_{\rm XXZ} & = & 
\sum \limits_{j}\left[J^{\perp}_{j}(s^{x}_{j}s^{x}_{j+1} +
s^{y}_{j}s^{y}_{j+1}) + J^{z}_{j}s^{z}_{j}s^{z}_{j+1}\right] \; ,
\label{H1}
\end{eqnarray}
where $\vec{s}_{j}$ are spin-1/2 operators at lattice sites $j$ separated
by spacing $a$, and
both $J^{\perp}_{j}$ and $J^{z}_{j}$ are random positive exchange
energies with a joint probability distribution $P_0(J^{\perp},J^{z})$.
Detailed information on the nature of the excitations in such
systems is encoded in the dynamical structure factor
$S^{\alpha \beta}(k,\omega)$ (where $\alpha \beta \equiv +-~
{\rm or}~ \alpha \beta \equiv zz$) with the $T=0$ spectral representation
\begin{equation}
S^{\alpha\beta}(k,\omega) =
\frac{1}{L}\sum \limits_{m}|\langle m|\sum
\limits_{j=1}^{L}e^{ikx_j}s^{\alpha}_j|0\rangle|^2 \delta(\omega - E_m) \; ,
\label{dynam1}
\end{equation}
where $s^{\pm}\equiv s^x \pm  i s^{y}$ and $\{ |m\rangle \}$ denote the exact
eigenstates of the system with energies $E_m$ ($|0\rangle$ is
the ground state, with $E_0 = 0$).

As mentioned earlier, we characterize transport in terms of the dynamical
conductivity $\sigma(\omega)$. The real part
$\sigma^{\prime}(\omega)$ of $\sigma(\omega)$
is defined by the relation $P(\omega)
= \sigma^{\prime}(\omega)|\nabla H|^2(\omega)$, where
$P(\omega)$ is the power absorbed per unit volume
by the system when a uniform magnetic field gradient $\nabla H(\omega)$
(where the field $H$ always points in the $z$ direction)
oscillating at frequency $\omega$ is applied along the length of
the chain.
From standard linear response theory, we have the following
Kubo formula for $\sigma^{\prime}(\omega)$ at $T=0$:
\begin{equation}
\sigma^{\prime}(\omega) = \frac{1}{\omega L}\sum \limits_{m}|\langle m|\sum
\limits_{j=1}^{L}
\tau_j|0\rangle|^2 \delta(\omega - E_m) \; ,
\label{kubo1}
\end{equation}
where $\tau_j =
-iJ^{\perp}_j(s^{+}_{j+1}s^{-}_{j}-s^{+}_{j}s^{-}_{j+1})/2$
is the current operator on link $j$ that transfers one unit of
the $z$ component of the spin from one site to the next, and the
frequency $\omega$ is taken positive for notational convenience.
Note that both $S^{\alpha \beta}(k,\omega)$ and
$\sigma^{\prime}(\omega)$, as defined here, are self-averaging
in the thermodynamic limit.

Randomness in the bonds is a relevant perturbation~\cite{doty} for
the pure XXZ chain when $0 \leq J^{z}/J^{\perp} \leq 1$; any amount
of disorder is thus expected to drive the system to strong disorder. In
this regime, the system can be treated by a strong-randomness
renormalization group (RG) that proceeds as follows~\cite{dasma,dsf,bhatt}:
We look for the bond
with the largest $J^{\perp}$
in the chain, say $J^{\perp}_{23}$
between spins $2$ and $3$ --- this sets the
energy cutoff $\Omega = \max \{ J^{\perp}_j \}$.
We first solve the corresponding two spin
problem (the neighbouring bonds are introduced later as perturbations).
So long as the $J^{z}$ are not large compared to the $J^{\perp}$,
the ground state of the two-spin problem will always be a singlet separated
by a large gap from the triplet excited states.
We can then trade
our original Hamiltonian in for another Hamiltonian
(determined perturbatively in the ratio of the neighbouring
bonds to the strongest bond)  which acts on
a truncated Hilbert space with the two sites connected by the
`strong' bond removed.
To leading order, this procedure
renormalizes the Hamiltonian ${\cal H}_{\rm 4sites} =
\sum_{j=1}^{3}J^{\perp}_{j}(s^{x}_js^{x}_{j+1}+s^{y}_js^{y}_{j+1})+
J^{z}_{j}s^{z}_js^{z}_{j+1}$ to
$\tilde{{\cal H}}_{14} = \tilde{J}^{\perp}_1(s^{x}_{1}s^{x}_4
+ s^{y}_1s^{y}_4) + \tilde{J}^{z}_{1}s^{z}_1s^{z}_4$
with $\tilde{J}^{\perp}_1 = 
J^{\perp}_{1}J^{\perp}_{3}/(J^{\perp}_{2} + J^{z}_{2})$
and $\tilde{J}^{z}_{1} = J^{z}_{1}J^{z}_{3}/2J^{\perp}_{2}$~\cite{dsf}.
Note that the new bond has length $\tilde{l}_1
= l_1 + l_2 + l_3$. This procedure, if it remains valid upon
iteration, thus
ultimately leads to a ground state of singlet pairs, with
pairs formed over long distances held together by
correspondingly weak bonds; this
is the RS state alluded to earlier.

A complete understanding of the possible states thus requires an
analysis of the effects of iterating the basic RG procedure. Such an
analysis was performed in Ref~\cite{dsf} leading to the following
conclusions:
If the $J^{z}$ dominate over the $J^{\perp}$,
this procedure rapidly becomes invalid and the ground state
actually has Ising Antiferromagnetic (IAF) order.
In all other cases, the
ground state is a Random Singlet state.
The low-energy (with energy
cutoff $\Omega \ll \Omega_c$, where $\Omega_c$ is the microscopic
cut-off) effective theory in these cases is written in terms of the
$n(\Omega)L$ `surviving' spin variables
(the bond-length $l$ between successive surviving sites is now a
random quantity).
When the $J^{\perp}$
dominate, this effective Hamiltonian has all $J^{z}_{j} = 0$,
and $J^{\perp}$ and $l$ drawn from a universal joint probability distribution
${\cal P}(
J^{\perp},l|\Omega)$
characteristic of the `XX Random Singlet fixed point' (XX-RS) of
the RG. Between the IAF phase and this XX-RS phase lie
two kinds of critical points. If the initial problem has full Heisenberg
symmetry ($J^{z} = J^{\perp}$ for each bond), the low-energy
effective Hamiltonian preserves this symmetry and has
bonds strengths and lengths drawn from the same probability
distribution ${\cal P}(J,l|\Omega)$. In the RG language, the Heisenberg
system is controlled by the XXX-RS fixed point.
Finally, in this language, the generic critical state is controlled
by the XXZC-RS fixed point---the low energy effective
theory has bonds and lengths drawn from a fixed point
distribution ${\cal P}_1(J^{\perp},
J^{z},l|\Omega)$ with the property
$\int dJ^{z}{\cal P}_1 = {\cal P}(J^{\perp},l|
\Omega)$. The probability distributions ${\cal P}$ and ${\cal P}_1$
become infinitely broad as $\Omega \rightarrow 0$; this implies that
the RG becomes asymptotically exact at low-energies, and in particular
predicts the ground state properties and low-temperature thermodynamics
correctly. Note that in all of the above, we have suppressed non-universal
scale factors multiplying the arguments of the functions $n$
and ${\cal P}$; these scaling functions however remain the same for all
systems that flow to any of the RS fixed points.

To use the foregoing for the calculation of dynamical or
transport properties, we need to explicitly keep track
of the renormalization of the operators that
enter spectral sums such as (\ref{dynam1}) and (\ref{kubo1}). 
We illustrate our general approach~\cite{us}
by first calculating the $T=0$ dynamical
conductivity at asymptotically low frequencies.
The first step is to work out
the rules that govern the renormalization of the current
operators.
Assume, once again, that $J^{\perp}_{23}$ is the
strongest bond. We
wish to work out perturbatively the renormalized operators 
$\tilde{\tau}_{1/2/3}$ that we trade in
$\tau_{1/2/3}$ for, when we freeze spins $2$ and
$3$ in their singlet ground state (the other current
operators to the left and right of this segment of our system
are left unchanged to leading order by the renormalization). Now,
note that these other operators have an
overall scale factor in them that is nothing but the corresponding $J^{\perp}$.
In order
to be consistent,
we clearly need to work out $\tilde{\tau}_{1/2/3}$ correct to $O(
\tilde{J}^{\perp}_{1})$ by adding the effects of virtual
fluctuations to the projection of $\tau_{1/2/3}$
into the singlet subspace~\cite{expopren}.
An explicit calculation~\cite{us}
gives the simple result: $2\tilde{\tau}_2 = 2\tilde{\tau}_{1/3}
= -i\tilde{J}^{\perp}_{1}(s^{+}_{4}s^{-}_{1} - s^{+}_{1}s^{-}_{4})$.
Thus, all three operators renormalize to the same operator, which
we will denote henceforth by $\tilde{\tau}_1$ for consistency of
notation.

As we carry out the RG and reduce the energy cutoff, the above
result implies that $\sum_{j}\tau_j$ renormalizes to
$\tilde{\sum_{j}}\tilde{l}_{j}\tilde{\tau}_{j}$, where $j$
now labels the $n_{\Omega}L$ sites of the renormalized lattice at cutoff scale
$\Omega$, and the $\tilde{l}_j$ are the
lengths of the renormalized bonds in this problem.
With this in hand, we run the RG until the
cutoff is reduced to $\Omega_{\omega} = \omega$ (note that
the operator $\tilde{\tau}_j$ linking two sites connected
by a strong bond $\{\tilde{J}^{\perp}_j,
\tilde{J}^{z}_j\}$
promotes the corresponding pair of spins from their singlet
ground state to the triplet state $|t_0\rangle$ with $m_z = 0$,
which is separated
from the ground state by precisely $\tilde{J}^{\perp}_{j}$) and
rewrite (\ref{kubo1}) as
\begin{equation}
\sigma^{\prime}(\omega) = \frac{1}{\omega L}
\tilde{\sum \limits_{m}}|\langle m|\tilde{\sum
\limits_{j}}
\tilde{l}_{j}\tilde{\tau}_j|0\rangle|^2 \delta(\omega -\tilde{E}_m) \; ,
\label{kuborg}
\end{equation}
where the tildes are a reminder of the fact that this spectral sum
now refers to the new Hamiltonian with cutoff $\omega$
which has only $n_{\omega} L$ sites (the couplings and
bond lengths in this problem are of course drawn from the
probability distribution characteristic of the fixed point to which
the system flows in the low energy limit).

We now have to calculate the spectral sum~(\ref{kuborg}) in this
new problem. The following crucial observation allows us to do this:
At the next step of the
RG, one would have looked for all the bonds in this renormalized problem that
have $J^{\perp}$ in the range $(\Omega_{\omega}, \Omega_{\omega} -d\Omega)$
and formed singlets out of the corresponding pairs of spins. 
The states with $\tilde{E}_m = \omega$ that
give the dominant contribution to the spectral sum (\ref{kuborg}) {\em
correspond precisely}~\cite{expdom} to promoting any one of these
pairs to the triplet state $|t_0\rangle$ under the action of
the current operator living on the corresponding bond. The matrix element
for this transition is just $\tilde{l}\omega/2$, where $\tilde{l}$ is
the length of the bond in question.
In the thermodynamic limit, we thus have
\begin{eqnarray}
\sigma^{\prime}(\omega) & \sim & \frac{n_{\Omega_{\omega}}}{\omega}
\int dl dJ^{\perp} \omega^2 l^2
{\cal P}(J^{\perp},l|\Omega_{\omega})
\delta(\omega - J^{\perp}) \nonumber \\
& \sim  & \ln(\Omega_c/\omega) \; ;
\end{eqnarray}
the last line is the leading behaviour for $\omega \ll \Omega_c$
obtained by using the results of Ref~\cite{dsf} for $n_{\Omega}$
and ${\cal P}$.
Notice that this analysis holds equally
well at all three RS fixed points.

Thus $\sigma^{\prime}(\omega)$ {\em diverges logarithmically}
for
small enough $\omega$ in the unusual `spin-metal' phase
controlled by the XX fixed point~\cite{lbmpaf} {\em as well as at
the critical point} separating this phase from
the insulating phase with
IAF order in the ground state. Note that this `metal-insulator' transition has
the curious feature that the quantum critical point separating
the conducting phase from the IAF insulator has the same $T=0$ transport
properties as the conducting phase.
On the insulating side,
$\sigma^{\prime}(\omega)$ is suppressed
below a pseudo-gap energy $E_g$. The dominant
contributions for $\omega \ll E_g$
come from Griffiths effects in which rare fluctuations
in the couplings of the Hamiltonian allow a long
finite segment of the system to be `locally' in the
`metallic' phase. A calculation of this
contribution gives~\cite{us}
a low-frequency conductivity $\sigma^{\prime}(\omega) \sim
\omega^{\alpha}\ln^2(\Omega_c/\omega)$, where $\alpha > 0$ is
a continuously varying, non-universal exponent that vanishes
at the transition.

The dynamical structure factor can be
calculated in a similar way. Consider first
$S^{zz}(k,\omega)$. One begins with the spectral
sum (\ref{dynam1}). The
leading order `operator
renormalizations' needed in this case are particularly
simple---each spin operator remains unchanged so long as it
is not part of a singlet and renormalizes to zero upon being
locked into a singlet state~\cite{expopren}. As before, we run the RG until
the cutoff $\Omega = \omega$ and do the spectral sum with
the renormalized operators in
the new problem. This renormalized sum may be evaluated by again
recognizing
that it is dominated~\cite{expdom}
by excitations to the triplet state $|t_0\rangle$
of pairs of spins connected by the renormalized bond
$\tilde{J^{\perp}} = \omega$.
The corresponding matrix element
is simply $(1-e^{ik\tilde{l}})/2$, where $\tilde{l}$ is
the length of the
strong bond (note that $\tilde{l}/a$ is an {\em odd} integer).
This allows us to write for $\omega \ll \Omega_c$ and $k=\pi/a +q$
\begin{eqnarray}
S^{zz}(k,\omega) & \sim & n_{\Omega_{\omega}}
\int dl dJ^{\perp}|1+e^{iql}|^2{\cal P}(J^{\perp},l|\omega)
\delta(\omega - J^{\perp})
\nonumber
\end{eqnarray}
at all three RS fixed points.
The calculation of $S^{+-}(k,\omega)$ is slightly
more involved as the gap to the relevant triplet excited state (with
$m^z =1$) of a pair of spins connected by a strong bond $(J^{\perp},J^{z})$
is now $(J^{\perp} + J^z)/2$. However, a careful analysis~\cite{us} gives
the same result as above for $\omega \ll \Omega_c$. 

Let us focus here on the
regime $|q|\equiv |k-\pi/a| \ll a^{-1}$ (in addition to
$\omega \ll \Omega_c$). In this regime, the integral can be
evaluated using
the results of Ref~\cite{dsf} for ${\cal P}$. This gives
the following rather unusual
{\em universal scaling form} at the
RS fixed points (i.e so long as the ground
state does not have IAF order)
\begin{eqnarray}
S^{\alpha \beta}(k,\omega) & = &
\frac{{\cal A}}{\omega \ln^3(\Omega_c/\omega)}
\Phi\left(|q|^{1/2}\ln(\Omega_c/\omega)/v_{\zeta}^{1/2}\right),
\end{eqnarray}
where $\alpha \beta \equiv + -~{\rm or}~zz$, ${\cal A}$ and $v_{\zeta}$ are
non-universal scale factors, and the fully {\em universal}
function
$\Phi(x)$ can be written as
\begin{eqnarray}
\Phi(x) & = &
\left(1+\frac{x(\cos(x)\sinh(x)+\sin(x)\cosh(x))}{
\cos^2(x)\sinh^2(x) + \sin^2(x)\cosh^2(x)}\right).
\end{eqnarray}

We now briefly summarize the
effects of weak (compared to the disorder)
enforced bond dimerization controlled by a dimensionless small parameter
$|\delta|$ that sets the difference between the probability distributions of
even and odd bonds~\cite{ising}.
The RG flows in the vicinity of the XX and XXX RS states
are known~\cite{yahybhgi,dsf};
the low energy properties are
controlled by lines of strong-disorder Random Dimer~\cite{yahybhgi}
fixed points ending
in the XX and XXX fixed points. These RD fixed points again describe ground
states that consist of singlet pairs; however, now the singlet
bonds preferentially start from an even (odd) site and end
at an odd (even) site for positive (negative) $\delta$.
Our approach readily
allows us to follow the full crossover of the low-frequency dynamical
conductivity and structure
factor from the XX and XXX-RS states to the
corresponding RD phases (only the non-universal prefactors differ in the two
cases). Here, we focus on results deep in the
RD phases, i.e for
frequencies such that
$\bar{\Gamma}_{\omega} \equiv |\delta|\ln(\Omega_c/\omega) \gg 1$.
For the conductivity, we obtain
$\sigma^{\prime}(\omega) \sim \ln(\Omega_c/\omega)
\bar{\Gamma}_{\omega}e^{-c_0\bar{\Gamma}_{\omega}}$, which
can be wriiten more explicitly as $\sigma^{\prime}(\omega) \sim
|\delta| \omega^{c_0 |\delta|} \ln^2(\Omega_c/\omega)$, where
$c_0$ is a non-universal constant. The RD phases are
thus seen to be gapless insulators. The dynamic structure factor
in the vicinity of $k=\pi/a$ can be written as
\begin{eqnarray}
S^{\alpha \beta}(k=\frac{\pi}{a} +q, \omega)
& = & \frac{{\cal C}|\delta|^3}{\omega^{1-\alpha}}
(1+\cos(c_1\bar{\Gamma}_{\omega}\bar{q})e^{-c_2\bar{\Gamma}_{\omega}\bar{q}^2}
),
\end{eqnarray}
where $\alpha = c_0 |\delta|$, $c_1$, $c_2$ and ${\cal C}$ are
non-universal constants, and
$\bar{q} \equiv qa/\delta^2$ is assumed $\ll 1$. This result has
striking oscillatory structure which is
best understood~\cite{us} as a novel signature of the sharply defined
geometry of the rare Griffiths regions that contribute to the scattering at
a given low frequency --- more precisely, the average length of the
relevant Griffiths
regions is of order $\bar{\Gamma}_{\omega}a/|\delta|^2$ while
the RMS fluctuations in the lengths are only of order
$\sqrt{\bar{\Gamma}_{\omega}}a/|\delta|^2$.

In the spin-1 case, the specific Hamiltonian we
consider is
\begin{equation}
{\cal H} = \sum \limits_{j} J_j \vec{S}_j \cdot \vec{S}_{j+1} \; ,
\end{equation}
where $\vec{S}_j$ are spin-1 operators and the $J_j$ are random, positive
bond-strengths; the corresponding distribution of $\ln(J_j)$ is
characterized by a width $W$.
A renormalization group analysis~\cite{Monthus}
reveals that the system flows to the analog of the XXX RS point described
earlier only when $W$ exceeds a critical value $W_c$. In this case,
our previous results for the dynamic structure factor and the dynamical
conductivity continue to apply. As $W$ is decreased, the system
undergoes a quantum phase transition to the so-called Gapless Haldane (GH)
phase~\cite{Monthus}; both the quantum critical point and the GH phase
in the vicinity of it are however still controlled
by strong-disorder fixed points~\cite{Monthus}.
Our technique may also be used to calculate
the dynamical conductivity at this critical point and in the
GH phase, although the details differ considerably from the
spin-1/2 case~\cite{us}. At the critical point, we find
\begin{equation}
\sigma^{\prime}(\omega)
\sim \ln^2(\Omega_c/\omega) \; ,
\end{equation}
which is a {\em stronger} divergence than in the
strong-disorder RS phase.
In the GH phase, the conductivity
goes as $\sigma^{\prime}(\omega) \sim \omega^{\alpha}\ln^2(\Omega_c/\omega)$
for log-frequencies $\ln(\Omega_c/
\omega)
\gg (W_c-W)^{-\nu/3}$ ($\nu$ is the correlation length exponent
known~\cite{Monthus}
to equal $6/(\sqrt{13}-1)$).
This phase is
thus a gapless insulator, not unlike the RD phase of spin-1/2
chains. The non-universal
exponent $\alpha$ is given as $\alpha \sim
(W_c -W)^{\nu/3}$.

We close with some remarks regarding experiments.
Our results for $S(k,\omega)$ near $k=\pi/a$ are clearly
of direct relevance to low-temperature INS experiments
on such spin chain systems in the regime where the
energy transfer $\omega$ satisfies $T \alt \omega \ll \Omega_c$.
Of particular interest would be an experimental
confirmation of the coherent oscillatory structure
predicted in
the RD phase. Regarding transport, we hope that our
results motivate experiments to
probe the spin conductivity in these systems.

We thank P.W. Anderson, R.N. Bhatt, F.D.M
Haldane, M. Hastings,  A. Madhav, R. Moessner, S. Sachdev, T. Senthil,
S.L. Sondhi, and A. Vishwanath for useful discussions. 
This work was supported by
NSF grants DMR-9809483 and DMR-9802468.


\begin{thebibliography}{99}
\bibitem{expdom} This is a consequence of the extremely broad
distributions of couplings in the low-energy effective theory.
\end{thebibliography}
\end{document}